\documentclass[12pt]{article}

\usepackage{amssymb,amsthm,amsfonts,psfig,epsfig}

\def\gappeq{\mathrel{\rlap {\raise.5ex\hbox{$>$}} {\lower.5ex\hbox{$\sim$}}}}
\def\lappeq{\mathrel{\rlap{\raise.5ex\hbox{$<$}} {\lower.5ex\hbox{$\sim$}}}}
\def\beq{\begin{equation}}
\def\eeq{\end{equation}}
\def\bea{\begin{eqnarray}}
\def\eea{\end{eqnarray}}
\def\bq{\begin{quote}}
\def\eq{\end{quote}}

 \def\ie{{i.e.}}
\def\eg{{\it e.g.}}

\def\i3{\mathbb{I}}

\def\eps{\epsilon}

\def\meg{\mu \rightarrow e \gamma}
\def\tmg{\tau \rightarrow \mu \gamma}
\def\teg{\tau \rightarrow e \gamma}

\begin{document}
\pagestyle{empty}

\begin{flushright}
ROMA1-TH/1397-04 \\ SACLAY-T05/08
\end{flushright}

\vskip 2cm

\def\thefootnote{\fnsymbol{footnote}}

\begin{center}
{\large \bf On Power and Complementarity of the Experimental Constraints 
on Seesaw Models } \vspace*{5mm}
\end{center}

\vspace*{5mm} \noindent \vskip 0.5cm

\centerline{\bf Isabella Masina$^{a)}$ and Carlos A. Savoy$^{b)}$} 

\vskip 0.5cm

\centerline{\em a) Centro Studi e Ricerche "E. Fermi", Rome, Italy, and}
\centerline{\em INFN, Sezione di Roma,  Rome, Italy}
\vskip 0.3 cm
\centerline{\em b) Service de Physique Th\'eorique, CEA - Saclay, France
\protect\footnote{Laboratoire de la Direction des Sciences de la
Mati\`ere du Commissariat \`a l'\'Energie
Atomique et Unit\'e de Recherche Associ\'ee au CNRS (URA 2306).}}

\vskip 2cm

\centerline{\bf Abstract}
We demonstrate the impact that present lepton flavour and CP violation data 
- neutrino oscillations, baryon asymmetry of the universe, 
flavour violations in charged lepton decays and lepton electric dipole moments - 
have on supersymmetric seesaw theories 
by analysing the class of models based on a $U(1)$ flavour symmetry.
The fermion $U(1)$ charges are determined to a large extent by the data,
the flavour patterns being \textit{naturally} defined through their choice. 
The selected models generically predict $\meg$ within the reach of the projected experiments,
which could be sensitive enough to exclude the whole class of them. 
By now, the present sensitivity to $\meg$ already provides stringent bounds 
on the contribution of the seesaw couplings to the lepton electric dipole moments.

\vspace*{1cm} \vskip .3cm  \vfill   \eject

\newpage

\setcounter{page}{1}
\pagestyle{plain}
\def\thefootnote{\arabic{footnote}}
\setcounter{footnote}{0}

\section{Introduction}\label{sec:INTRO}

The simplest way to provide for light neutrino masses is the seesaw
mechanism \cite{seesaw}. However, even in its basic type I version, 
the flavour pattern of the Dirac couplings $Y_\nu$ and the Majorana mass 
matrix $M_R$ remains largely undetermined. Indeed, the already very accurate 
data on the atmospheric and solar mass  differences and mixing angles 
cannot discriminate  between effective neutrino mass matrices, $m_\nu$, with 
different overall patterns. The potential measurements of other parameters 
in $m_\nu$ cannot resolve the indeterminacy  which is an intrinsic feature 
of the heavy neutrino decoupling process - see e.g. \cite{CDI}-\cite{Petcovetal}. 
Complementary informations on the $Y_\nu$ and $M_R$ flavour patterns  
are to be found in the quantum effects  involving the heavy neutrinos. 
It is well known that this is the case of the {\it lepton asymmetry of the 
universe} \cite{FukYan} and, assuming supersymmetry, of {\it charged 
lepton dipole transitions}, like the lepton flavour violating (LFV) decays 
$\ell_i \rightarrow \ell_j \gamma$ \cite{BorMas} and the CP violating 
(CPV) lepton electric dipole moments (EDM) $d_i$ \cite{EDM-Ellis, 
EDM-Isa}.  Without supersymmetry the large LFV discovered in neutrino 
oscillations and the CPV needed for leptogenesis do not lead to any 
measurable effect in the charged lepton interactions \cite{LFV-SM}, in
spite of the high sensitivity reached, in particular, in 
$\mu \rightarrow e\, \gamma$ \cite{exp-mueg-p} and $d_e$ 
\cite{exp-de-p}. Anyhow, protecting the Fermi scale with respect to 
the large mass scales involved in seesaw models requires a suitable mechanism  
- such as the supersymmetry framework, which is adopted in this work - 
and the corresponding new physics at scales not far from the TeV region. This
is basically the region that lepton experiments are already testing for LFV 
and CPV sources.

The predictions for leptogenesis and charged lepton dipole transitions
have very different dependences on $Y_\nu$ and $M_R$ with respect to
neutrino oscillations, yet the extraction of the constraints on the seesaw 
parameters is not straightforward. Firstly, there are  more parameters 
than experimental constraints; secondly,  their analysis relies on additional 
assumptions, \eg, the neutrino initial abundance and the supersymmetric 
masses. Therefore one is forced to reduce the number of parameters by 
considering a model or a class of models at a time.
Of course, the interest of these analyses depends on the naturalness of the assumptions and the 
robustness of the results with respect to variations in the parameters.  

In this paper, we contribute to this exploratory work and show the power 
and the complementarity of the combined constraints from all these 
experiments by studying a popular class of models: the supersymmetric 
Froggatt-Nielsen ones where the seesaw parameters are restricted by a 
$U(1)_F$ flavour symmetry \cite{FN}.  Actually, these models 
\cite{Ramond, Vissani, SatoYana} seem to  best balance theoretical simplicity 
and  consistency with experiments  \footnote{See \eg ~\cite{AFM3} 
for a comparison with less or more structured $U(1)_F$ models.}.  Their 
basic ingredients are: the charges of the three lepton doublets, $\ell_i$ 
($i=1,2,3$), and those of the three heavy right handed neutrinos, $n_{i}$;
one small parameter, $\eps$, associated to the breaking of $U(1)_F$ and 
to a charge $-1$. The abelian flavour symmetry fixes the matrices $Y_\nu$ 
and $M_R$  up to $O(1)$ coefficients and thus the $\eps$ dependence of 
the various lepton observables in terms of the charges $\ell_i$ and $n_{i}$. 
This class of models has at least seven attractive features : 
\textit{(i)} there is only one small parameter $\eps$ besides the $B-L$ 
breaking scale $v_{B-L}$;  \textit{(ii)} the hierarchies in $M_R$ depend 
only on $n_{i}$; \textit{(iii)} the ratio between the (non-vanishing)  $Y_\nu$ 
couplings of any right handed neutrino  only depends on corresponding
$\ell_i$ differences, so , \textit{(iv)} $m_\nu$ has  hierarchical eigenvalues 
depending only on $\ell_i$ differences; \textit{(v)} the models are
natural and the hierarchies do not rely on the unknown $O(1)$ coefficients; 
\textit{(vi)} the texture (the zero entries) of $Y_\nu$ is determined by 
supersymmetry (analytic superpotential); \textit{(vii)} the hierarchies in 
the $Y_\nu$ elements are fixed by the $U(1)_F$ invariance of the 
superpotential. It has one (well known) slightly unattractive feature: some amount of tuning 
is needed to preserve the hierarchy between the solar and the atmospheric
mass differences.

Of course, many aspects of these models have been studied in the literature, 
mostly at a time when the experimental situation of neutrino oscillations was 
more poorly established, leaving more freedom for the parameters $\ell_i$
and $\eps$. One of the first studies on leptogenesis considered models of this 
kind and showed that the final lepton asymmetry does not depend on $M_2$
and $M_3$ but only on the lightest eigenvalue, $M_1$, of $M_{R}$ 
\cite{BuchYana}. LFV has been considered in some models of 
this class, and it was noticed that with $O(1)$ Yukawa couplings in $Y_\nu$ 
the predictions are close to the experimental limits for $\meg$ 
\cite{BuchDelViss}-\cite{LMS1}.  The significant improvement in 
our knowledge of neutrino oscillation parameters motivates our reappraisal 
of the abelian flavour symmetry models. 

We show that, in this framework, the mass eigenvalues $M_{i}$ of 
$M_{R}$ are already almost fixed by present experiments and 
basically the whole class of models might be excluded by future 
ones \cite{exp-mueg-f}. In addition,  we also derive bounds on the lepton EDM from 
the limits on $\meg$ (which do  not apply to other seesaw models) 
to conclude that the phases in $Y_\nu$ alone are not enough to 
radiatively induce lepton EDM even at the level of the planned  
experiments \cite{exp-de-f, exp-dmu-f}.
 
We first consider models where, from the choice of the  $U(1)_F$ charges,
no texture zeros are natural  in $Y_\nu$, in the sense that they are not 
required by supersymmetry analyticity. A characteristic of this whole class
of models is that, in the seesaw formula, $m_{atm}$ gets roughly equivalent
contributions from \textit{all} heavy neutrinos. This fact by itself 
puts an upper bound $O(5\times 10^{14})$GeV on all the $M_{i}$. 
Oscillation experiments then  ask for $\ell_2=\ell_3=\ell_1-1$ and 
$\eps = O( m_{sol}/m_{atm} ) $ (actually close to the Cabibbo angle, 
often chosen for $\eps$ in models for the quark mass hierarchies). The 
models then yield the  prediction $U_{e3}=O(m_{sol}/m_{atm})$, to be 
tested in the near future. The heavy neutrino charges, $n_{i}$ remain free 
and so do the $M_{i}$. 

The next step is to show that the interplay between the constraints from 
leptogenesis, LFV and EDM, allow to gain significant information on the 
spectrum of right handed neutrinos in this class of models. With the 
$m_{atm}$ matrix elements fixed up to $O(1)$ coefficients, leptogenesis 
is possible in these models only if $M_{1} = O(10^{11})$GeV. The rate 
of LFV depends linearly on $M_3$, while EDM depend on the product 
$M_2 M_3$.  The limit on $M_3$ from the $\meg$ search presently is two 
orders of magnitude above the leptogenesis value for $M_{1}$ and is
on the way to be improved. Thus, this class of models has the nice 
feature that the comparison between theoretical parameters and the more 
restrictive experimental data is  one-to-one in practice, giving also some 
predictions for quantities to be better measured in the future. As a matter 
of fact, EDM turn out to be so small that the only constraint on $M_{2}$ 
comes from its definition, $M_{1} \leq M_{2} \leq M_{3}$. This is because
there is a relation between lepton EDM and LFV lepton decays  while the 
latter are related by the $U(1)_{F}$ symmetry. Therefore all these 
quantities are bounded from the strong limits on $\meg$, providing
further tests of the models.

In this class of models, the strong upper limits on the heavy neutrino 
masses are related to the fact that their contributions to $m_{atm}$ are comparable. 
However, the heaviest neutrino can be \textit{naturally} decoupled from the
$\mu$ and $\tau$ doublets if one chooses $n_{3}$ low enough
so that the corresponding matrix elements in $Y_{\nu}$ are
zero by the analyticity of the supersymmetric Yukawa couplings.
This alternative class of models is also studied here and we show that 
$M_{3}$ gets close to $O(M_{GUT})$. However, the somewhat surprising
result is that the would-be decoupled heavy neutrinos are even more
coupled in $Y_{\nu}$! As a consequence these models are already
generically excluded by the $\meg$ data. Barring a
complete decoupling of any heavy neutrino from all light ones
is enough to select a very restricted, hence predictive, class of
models when the experimental constraints are imposed.

A comment is in order on the supersymmetric flavour problem in the context of flavour theories, 
in particular those based on abelian groups. In general, the breaking of the $U(1)_{F}$ symmetry 
defining the small number $\eps$ induces supersymmetry breaking components for both $D$ and $F$ 
auxiliary fields along $U(1)_{F}$ charged directions, which in turn generically generate soft
terms with LFV misalignment and CPV phases \cite{dudas}. 
The flavour non-diagonal contributions to the soft slepton masses come out of order 
$(\ell_j - \ell_i ) \eps^{\ell_j - \ell_i }$ and $(n_j - n_i ) \eps^{n_j - n_i }$ 
with respect to the diagonal slepton masses \footnote{The origin of these flavour 
problems is related to the structure of the Kahler metric and the Yukawa
interactions discussed in Appendix B. When the $U(1)_{F}$ is anomalous, 
its  breaking is induced by a Fayet-Iliopoulos term and the associated $D$-terms 
give rise to an inversed hierarchy for the scalar masses and potentially large LFV and CPV 
in the mass matrices in the charged lepton flavour basis \cite{dudas}.}. 
They are analogous (up to charge difference factors) to the pattern of the radiative 
corrections discussed in this paper, since their $\eps $ dependence is basically 
fixed by the $U(1)_{F}$ broken symmetry too.
A fine-tuning between the mostly constrained tree-level and radiative 
contributions looks very unnatural. 
Hence, one can naturally assume that each contribution is separetely restricted and should 
be separetely cared for. In particular, a suppression mechanism for both the $D$ and $F$ 
supersymmetry breaking associated to the $U(1)_F$ symmetry breaking, would
lessen the tree-level contributions without any impact on the radiative ones.  

The paper is organised as follows. The models are defined in section 2.
The resulting effective neutrino mass matrix is confronted to the
oscillation data constraints in section 3 and the general structure of
$m_{\nu}$ is obtained. In section 4, the leptogenesis requirements
are imposed on the selected models to obtain the order of magnitude
of $M_{1}$.  The seesaw induced   LFV effects in supersymmetric
models are briefly reviewed in section 5, with the limits from the
BR($\meg$)  upper bound given as functions of the slepton and
gaugino masses. In the case of the models without holomorphic zeros
in  $Y_{\nu}$, one gets upper limits on $M_{3}$ which are quite
low and will be further lowered by the future $\meg$ measurements. 
Section 6 starts with a recall of the general seesaw induced EDM formulae 
written so to make their connection to the LFV transitions manifest. Some 
details are given in appendix A. Then,  they are applied to the models 
discussed here where the indirect limits on $d_{e}$ from $\meg$ supersede
those from the $d_{e}$ experimental searches. In section 7, models
with naturallly vanishing couplings in  $Y_{\nu}$ are considered and
analysed along the same lines as in the previous sections. Some details
are given in appendix B.


\section{The model} {\label{sec:MODEL} 

The model is defined in the lepton sector by the following set of
$U(1)_F$ charges:  $\ell_i$, $e_i$ and $n_i$, $(i=1,2,3)$ for the
lepton electroweak doublets, charged and neutral singlets,
respectively; $h_u$, $h_d$ for the electroweak symmetry breaking Higgs
fields with v.e.v.'s  $v_u$, $v_d$, respectively; $2q$ for the field(s)
breaking lepton number by two units in the Majorana masses  and
associated with the scale $v_{B-L}$. The $U(1)_F$ symmetry breaking
field charge is normalized to  $-1$ and the small flavour symmetry breaking
parameter, $\eps$,  is provided by the ratio of its v.e.v. and the cut-off of
the flavour theory.  

We first concentrate on the class of models such 
that all entries in the leptonic mass matrices are non-zero in the $U(1)_F$
basis, as any charge imbalance in the corresponding terms in the
effective superpotential can be compensated by analyticity-preserving
insertions of $\eps$.  Hence, the flavour structure of the matrices in
the leptonic sector is determined, up to $O(1)$ coefficients, by the
following non-negative powers of $\eps$:

\beq
(Y_{\ell})_{ij} = g_{ij} \ \eps ^{e_i + \ell _j + h_d} \, ,\quad
(Y_{\nu})_{ij} = h_{ij}\  \eps ^{n_i + \ell _j + h_u} \, ,\quad
(M_R)_{ij} = r_{ij}~ \eps ^{n_i + n _j + 2q} \, v_{B-L}\ , 
\label{model}
\eeq

\noindent where all the elements of the matrices $g$, $h$ and $r$ are $O(1)$.
The effective neutrino mass matrix results in the expression

\beq
(m_{\nu})_{ij} = (Y_{\nu}^T M^{-1}_R Y_{\nu})_{ij} v^2 _u =
(h^T r^{-1} h)_{ij} \eps ^{\ell _i + \ell _j  +2 h_u - 2 q} 
\frac{v^2 _u}{v_{B-L}} ~. 
\label{m_nu}
\eeq

\noindent After studying these models in detail we shall extend our analysis to
models where some powers in (\ref{model}) are negative and the
corresponding entry vanishes in the $U(1)_F$ basis.

Models with non-zero entries have several interesting properties 
for the purposes of our analysis.  
Firstly, integrating out the Froggatt-Nielsen 
field preserves the analyticity in the effective theory, at least at the lowest
order in $\eps$ and the powers of $\eps$ are given by the charge 
imbalance in the effective operators. This is evident in (\ref{m_nu}), 
where $m_\nu$ is independent of the charges $n_i$ \cite{Ramond} and 
gives no information on the heavy right handed neutrino spectrum.  
Secondly, the structure -- in terms of powers of $\eps $ -- of the unitary
transformations diagonalizing the Yukawa couplings and the Majorana
mass matrix in (\ref{model}) is fixed by the differences between the
$U(1)$ charges of different generations.  As a consequence, in the
basis where $Y_\ell$ and $M_R$ are diagonal, only the elements of $g$,
$r$ (now diagonal) and $h$ change with respect to eq. (\ref{model}),
but they generically remain $O(1)$ (for a detailed proof see \eg \cite{BuchYana}).  
Since the abelian flavour symmetry only fixes the
powers of $\eps$ up to arbitrary $O(1)$ coefficients, it is
convenient and generically equivalent to work directly in the basis
where $Y_\ell$ and  $M_R$ are diagonal.  Hence, in the following we
choose this basis while keeping the same notations as above for
simplicity.  In particular (\ref{m_nu}) is now in the physical charged
lepton basis used in the fits of neutrino oscillations, also convenient to study 
charged LFV and CPV transitions, of course.

The heavy right handed neutrinos possess a characteristic property in
this class of models: the branch ratios for their decays into each light 
lepton flavour are all of the same order of magnitude and proportional to 
$\eps ^{2\ell _j}$. This property is partially lost when some matrix 
elements in (\ref{model}) vanish by analiticity as discussed 
in section \ref{sec:DECOUPLING}.

\section{Neutrino Oscillations} \label{sec:OSC}

Let us start with the comparison of the effective neutrino mass matrix
in (\ref{m_nu}) and the corresponding parameters as determined by the
experiments. In our analysis we adopt the $3-\sigma$ ranges of ref. \cite{fit}:

\bea
m_@ = \sqrt{m_3^2 - m_2^2} = (.05 \pm .01 ) ~\mathrm{eV} ~~~,~~~
m_\odot = \sqrt{m_2^2 - m_1^2} = (.0091 \pm .0005 ) ~\mathrm{eV}~
\nonumber\\
.71 \le \tan \theta_{23} \le 1.45 ~~~,~~~\tan \theta_{12} = .65 \pm .12~~~,
~~~\tan \theta_{13} \le .21~~.~~~~~
\label{nuosc}
\eea

\noindent Of course, this comparison is present in several papers, 
\eg~\cite{Ramond, Vissani, SatoYana}, but the recent improvements in the
experimental determination of neutrino oscillation parameters - in
particular in the solar sector - now allow for a basically unique
solution, that we turn to discuss.

First of all, the very large atmospheric mixing angle requires 
$\ell_2 = \ell_3 = \ell $. It is convenient to introduce the parameter $\rho =
\Delta_{23} /m^2_3 $, where $\Delta_{23}$ is the sub-derminant 
of the $i,j=2,3$ sub-matrix of $m_\nu$ and $m_3$ is its largest eigenvalue. 
The bound on

\beq
\sin \theta_{13} \approx \eps ^{\ell _1 - \ell } ~\frac{\rho +1}{\sqrt{2}}
\frac{|(h^T r^{-1} h)_{12} + (h^T r^{-1} h)_{13}|}
{|(h^T r^{-1} h)_{22} + (h^T r^{-1} h)_{33}|} 
\label{theta_13}
\eeq

\noindent requires $ \ell_1 > \ell $ leaving the larger elements
in the $i,j=2,3$ sub-matrix of eq. (\ref{m_nu}) to be of $O(m_@)$.
We can take $ \ell_1 = \ell + 1$ up to a trivial re-scaling of $\eps$.
From these two requirements, the flavour structure of $m_\nu$ in this
class of models is fixed :

\beq
(m_{\nu})_{ij}  = ~ (h^T r^{-1} h)_{ij}~\eps^{\delta_{i1}
+\delta_{j1}}~~\eps ^{2(\ell +h_u - q)}  \frac{v^2 _u}{ v_{B-L}}~~~~.
\label{m_fit}
\eeq

This structure has also to accomodate the ratio $m_\odot / m_@ = 0.18 \pm 0.05$ 
as well as the relatively large but not maximal value of $\tan \theta_{12}$. 
One has $\rho \approx  \cos^2 \theta_{12} m_2/ m_3$ and  
$\tan \theta_{12} = O(\eps / \rho ) $.
Then, for $m_@ \approx m_3$ and $m_\odot \approx m_2$,
by using the experimental range for the solar angle one obtains  
$O(\eps) = m_\odot \sin 2 \theta_{12} / 2 m_@ \approx .08 \pm .01$.  
In the spirit of the whole approach, we take advantage of the 
$O(1)$ arbitrariness in the matrices $h$ and $r$ to fix  $\eps$ at the central 
value of $m_\odot / m_@$, $\eps= 0.18$, from now on. 
From (\ref{theta_13}), it follows that  $U_{e3} = O(\eps)$, 
not so far from the present upper bound \cite{SatoYana,AFM3}.

Clearly, $\rho = O(\eps )$ requires a relatively mild tuning of the
$O(1)$ coefficients, inherent to this class of models \cite{Ramond},
which remains a weak point of the present class of models
\footnote{Models with single right handed neutrino dominance, 
where the suppression of $\rho$ becomes natural, were specifically 
proposed in order to overcome this problem \cite{King}.}. 
To illustrate it, in fig. \ref{F-distr}, the $\rho$ distribution is shown 
for randomly generated $O(1)$ complex matrix elements in $h$ and $r$ 
and also for the small subset of the generated models that fulfil {\it all} 
the experimental constraints (\ref{nuosc}) - the latter magnified by a
factor of $10$ in the figure. The experimentally allowed region of
$\cos^2 \theta_{12} m_\odot/ m_@$ is also shown for comparison.
One can estimate from the figure the probability for the generated
models to satisfy the required tuning in $\rho$. Of course, the
relatively high precision of the neutrino oscillation data
(\ref{nuosc}) further reduces the experimentally viable models to the
small fraction shown in the figure.
  
\begin{figure}[!h]
\vskip 0.5 cm
\centerline{
\psfig{file=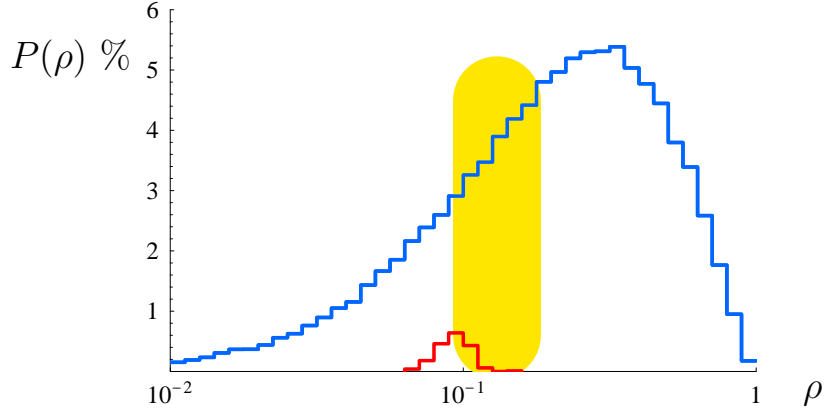,width=.6 \textwidth }
\put(-300,130){ \large $P(\rho)$ \% }
\put(0,5 ){ \large $\rho$ }  }
\caption{Distribution of $\rho$ obtained from a sample of models with
charges $\ell_1-1 = \ell_2 = \ell_3\equiv \ell$, $\eps=0.18$ and matrix
elements $r_i $, $h_{ij}$ randomly generated among the complex numbers
with absolute values between $0.3 - 3$. The upper (blue)
distribution refers to the whole sample, while the lower (red)
distribution is obtained by keeping only the models in the sample
satisfying all the present neutrino oscillation bounds (\ref{nuosc}). The
latter has been magnified by a factor of $10$ for graphical reasons.
The (yellow) shaded region represents the present range of
$ \cos^2 \theta_{12}m_{\odot} /m_@ $. }
\label{F-distr}
\vskip 0.5 cm
\end{figure}

The heavy Majorana neutrinos all contribute the same order of magnitude
to each element of $m_\nu$ independently of their masses, $M_i$.
However, there is an upper bound on their masses if their couplings
$Y_{\nu}$ are to remain perturbative. Indeed, from eq. (\ref{model})
one has

\beq
\label{specif}
\eps ^{2(\ell +h_u + n_i)} ~r_i~ \frac{v^2 _u}{M_i} 
= \eps ^{2(\ell +h_u - q)}  ~\frac{v^2 _u}{ v_{B-L}}
= \frac{ m_@}{ d_@ }  ~~~~~~,~~\forall i
\eeq

\noindent where $d_@$ = O(1) is the analog of $m_@$ for the
dimensionless matrix $(h^T r^{-1}
h)_{ij}~\eps^{\delta_{i1}+\delta_{j1}}$. Since by assumption $\ell +h_2
+ n_i \ge 0$, we obtain the upper bound for each $M_i$

\beq
\frac{M_i}{M_@} = \eps ^{2(\ell +h_u + n_i)} d_@ r_i \lesssim  O(1)~~~~~~,~~~~~
M_@ = \frac{v^2 _u}{ m_@} \simeq 5 \times 10^{14}~ \mathrm{GeV}~~~~.
\label{Mat}
\eeq

Another characteristic property following from $\ell_1 = \ell +1$,
relevant for the predictions of the next sections, is the
fact that {\sl all} the right-handed neutrinos,  $\nu ^c _i$, couple to
the first family doublet $ \ell_1$ by a factor of $\eps$ less than to
the other two family doublets.  This means that $e$, $\nu  _e$ are
expected to be less important for leptogenesis (the decays into $\mu$,
$\nu_\mu$ and  $\tau$, $\nu_\tau$ are dominant) and also that LFV and
CPV transition amplitudes involving the electron are roughly reduced by
corresponding factors of $\eps = m_\odot / m_@ $. 

We now turn to discuss how other experimental data can give
informations on the flavour structure of right handed neutrinos,
starting with the constraints from the assumption of leptogenesis.

\section{ Leptogenesis} \label{sec:LG} 

One of the very first explicit calculations of leptogenesis with the
seesaw \cite{BuchYana} mechanism adopted models of the same class 
discussed here.
It found that the lepton asymmetry critically depends on $M_1$ and is 
quite independent of the right handed neutrino mass hierarchy.  Of course, 
the neutrino oscillation parameters were less constrained at that time.
Here,  we update and slightly generalize it and we explicitely find the
favoured range of $M_1$, which is used in the next sections.

In the notations of ref. \cite{bdp}, adapted to the supersymmetric case
\cite{Covi, Giudice, DiBari}, the resulting expression for the baryon asymmetry of the
universe is:

\bea
\eta_B = - 10^{-2} \varepsilon_1 \kappa_f ~~,~~~~~~~~~
\varepsilon _1 = \frac{1}{8 \pi} \sum_{j \neq 1}
\frac{\mathrm{Im}[(Y_\nu Y_\nu^\dagger)_{j1}^2]}{(Y_\nu Y_\nu^\dagger)_{11}}
g\left(\frac{M_j^2}{M_1^2}\right) ~~,~~~\nonumber \\
g(x) = - \sqrt{x} \left[ \frac{2}{x-1} + ~\ln \frac{1+x}{x} \right]
~\stackrel{\footnotesize{x >>1}}{\longrightarrow} -\frac{3}{\sqrt{x}}
~~,~~~~~~~~~~~~~~~ \label{etaB}
\eea

\noindent where $\kappa_f$ is the efficiency factor. In the present class 
of models we can take advantage of the $x >>1$ limit because both
conditions that could invalidate this approximation 
are unnatural. Indeed, without fine-tuning, $M_1$ and $M_2$
cannot be degenerate at the level required for a resonant enhancement
\cite{pilaf}, and $M_2$ and $M_3$ cannot be a pseudo-Dirac pair \cite{Raidal}.  
In the strong washout regime, where the effective neutrino mass $\tilde
m_1 =(Y_\nu Y_\nu ^\dagger)_{11} v_u^2/M_1  $ is larger than the
equilibrium neutrino mass $m_* \approx 0.8 \times 10^{-3}$eV \cite{DiBari}, the
dependence of $\kappa_f$ on the initial abundance is very small and
well approximated by the power law \cite{bdp, Giudice, DiBari}:  $\kappa_f
\approx (1.5 \pm 0.7) \times 10^{-4} ~\mathrm{eV} /\tilde m_1$.  Thus,
successful leptogenesis yields a general lower limit on $M_1$ as a
function of $\tilde m_1$, as displayed, \eg , in ref. \cite{bdp, DiBari}.

In the models considered here, defining the $O(1)$ matrix
$H_{ij} = \sum_k h_{ik} h_{jk}^* \eps^{2\delta_{k1}}$, one gets:

\beq
\varepsilon_1  =  \frac{3}{8 \pi} \frac{M_1 m_@}{v^2 _u}
\sum_{j \neq 1} \frac{ \mathrm{Im}[H_{j1}^2] }{ H_{11} r_j d_@ }
= \frac{3}{8 \pi} O(\frac{M_1 }{M_@ } )
~~,~~~~~~ \tilde m_1 = m_@ \frac{H_{11}}{r_1 d_@} = O(m_@ )~.~~~
\label{eps}
\eeq

\noindent Hence the approximation for $\kappa_f$ in the strong washout 
regime is valid and one gets for the baryon asymmetry of the universe:

\beq
\eta_B  \approx  \frac{2 \pm 1}{2}  \frac{10^{-7} \mathrm{eV}}{m_@} \frac{M_1}{M_@}
\sum_{j \neq 1} \frac{\mathrm{Im}[H_{j1}^2]}{H_{11}^2} \frac{r_1}{r_j}  ~~~.
\label{constr}
\eeq

\noindent to be compared to the experimental value 
$\eta_B = (6.3 \pm 0.3) \times 10^{-10} $ \cite{BAU}. 
Therefore one derives, within the uncertainty of a factor $1/2 - 2$, 
the value of $M_1/M_@$ required by leptogenesis. This is shown in 
fig. \ref{F-leptog} for the randomly generated models selected by the 
neutrino oscillation data as discussed in section \ref{sec:OSC}. The results 
favour $M_1 = O( 10^{11} $ GeV).  It is well known that leptogenesis at
such scale hides a potential gravitino problem which is solved if the
gravitino is either stable \cite{Boltz} or heavier than $O(2$ TeV)
\cite{Kawa}. Notice that, since $\varepsilon _1 $ in eq. (\ref{eps})
turns out to be close to the DI upper bound \cite{dibound}, the above
prediction for $M_1$ is actually close to its general lower limit for
$\tilde m_1 = O(m_@ )$ \cite{bdp}.

\begin{figure}[!t]
\vskip .5 cm
\centerline{  \psfig{file=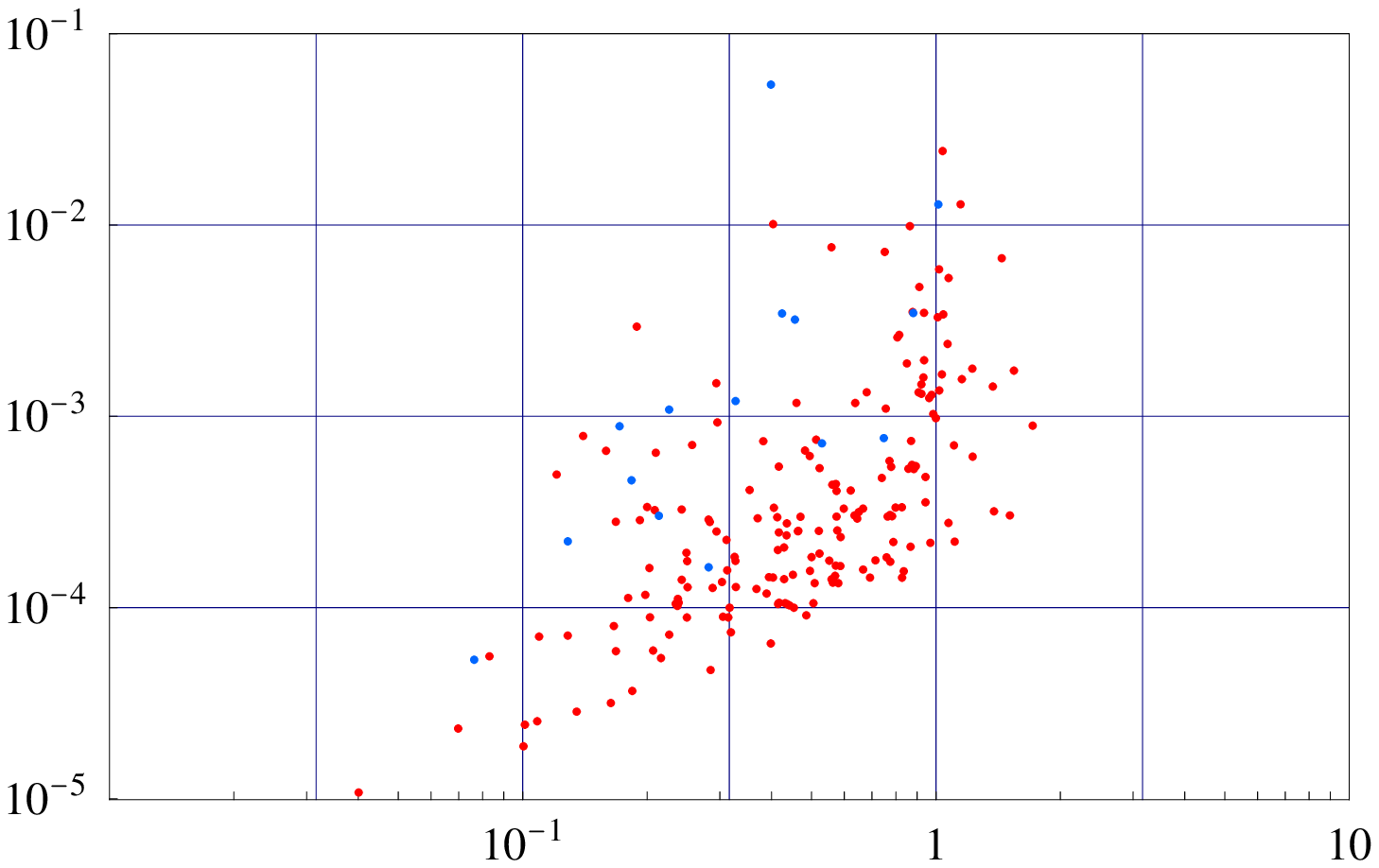,width=.7 \textwidth}
\put(-75,-5){ \large $\tilde m_1 / m_@$}
\put(-350,150){ \large $M_1 / M_@$}  }
\caption{Scatter plot in the plane $M_1/M_@$, eq. (\ref{constr}), and
$\tilde m_1 /m_@$, eq. (\ref{eps}), for the models (described in the
previous section) satisfying all the neutrino oscillation experimental
data.  The value of $M_1/M_@$ brings a theoretical uncertainty of a
factor $1/2 -2$. For the points in red the sine of the leptogenesis phase is 
$ \geq 0.1$ while for those in blue it is $ < 0.1$. }
\label{F-leptog}
\vskip .5cm
\end{figure}

Notice also that in $H_{j1}$ the electron term is reduced by $\eps ^2$ as
compared to the $\mu $ and $\tau$ ones, so that the leptogenesis phase
should mostly result from the latter.

\section{ LFV } \label{sec:LFV}

We now turn to the experimental data that can provide informations on the 
heaviest right handed neutrino, namely, the LFV decays of the charged 
leptons. In a theory based on broken supersymmetry,  $\ell_i 
\rightarrow \ell_j \gamma$ arise from the possible misalignment
between the slepton and the charged lepton mass matrices.  The
expansion of the transition amplitude in terms of the relative
misalignment of the slepton masses - or insertion approximation - is
justified by the fact that these LFV rates are expected to be
relatively small, as attested by the bounds on $\meg$.  There is a rich
literature on the experimental constraints on this misalignment that
emphasize the importance of the supersymmetric flavour problem in the
lepton sector \cite{liter, hisano}. Here,  we make use of the results in 
\cite{sleptonarium}.

In supersymmetric theories the radiative corrections are wave function
renormalisations by hermitean matrices, whose flavour and CP structures
depend on the Yukawa couplings.  In the context of the supersymmetric
seesaw models, the radiatively induced misalignment was emphasized a
long time ago \cite{BorMas} and more recent analyses 
\cite{CDI, BuchDelViss, SatoTobe, LMS1, LFVvari, LFVegm2} 
were prompted by the determination of the neutrino oscillation parameters.  
Starting from CP and flavour conserving soft terms and taking $M_R$ and
$Y_{\ell}$ real and diagonal at $M_\mathrm{Pl}$, the flavour and CP
violations in slepton masses are introduced in the RG equations only by
heavy neutrino $\nu^c_k$ loops through their flavour changing
couplings $Y^\dagger_{\nu ik} Y_{\nu kj}$, $i \neq j $, until they
successively decouple at $M_k$ yielding corresponding factors $\ln
(M_{\mathrm{Pl}}/M_k )$.  Since $Y_{\nu}$ couples $\nu^c_k$ to the
lepton doublets, the seesaw radiative corrections mainly affect the
doublet slepton masses $m^2_{L ij}$.  Assuming universality in soft
masses at $M_{Pl}$, the misalignment in $m^2_{L}$ at the lowest order
in the couplings $Y_{\nu}$ (defined at $M_{Pl}$) is \cite{BorMas}:

\bea
\label{Cgen}
\delta^{LL}_{ij} & = & \frac{m^2_{L ij}}{\bar m^2_L} = - \frac{1}
{(4 \pi)^2} \frac{6 m_0^2 + 2 a_0^2}{\bar m^2_L} C_{ij} ~, \nonumber \\
C_{ij} & = & \sum_{k} C^k_{ij} = \sum_{k} Y^*_{\nu ki} Y_{\nu kj} 
\ln \frac{ M_\mathrm{Pl} }{ M_k } ~ 
\eea

\noindent where $m_0$ and $a_0$ are the universal soft breaking
bilinear and trilinear terms at $M_{Pl}$, respectively, and $\bar m_L$ is the
average doublet slepton mass.  We have isolated the contribution from
each $\nu ^c _k$ state, $C^k_{ij}$, for later convenience.

In the basis where $Y_{\ell}$ and $M_R$ are diagonal, the mass insertion
approximation reads

\beq
\label{BR}
\mathrm{BR}(\ell_i \rightarrow \ell_j \gamma) =  
10^{-5} ~\mathrm{BR}(\ell_i \rightarrow \ell_j \bar \nu_j \nu_i)
~\frac{M^4_W}{\bar m^4_L} \tan^2 \beta |\delta^{LL}_{ij}|^2 F_{susy}
\eeq

\noindent where $F_{susy}=O(1)$ is a function of supersymmetric masses which
includes both the chargino and neutralino exchange (see \eg ~\cite{sleptonarium}, 
and references therein).  With eq. (\ref{Cgen}),
the resulting general limits on slepton misalignment
\cite{sleptonarium} translate into limits on the $|C_{ij}|$ as
functions of the supersymmetric masses and $\tan\beta$ \cite{LMS1, IMsusy02} 
\footnote{The figures in \cite{LMS1} only take into account
the chargino contribution, not dominant in some regions of the
supersymmetric masses \cite{sleptonarium}. The neutralino contributions
are included in the plots of ref.  \cite{IMsusy02}.}.  
In mSugra one can express the dependence on the susy spectrum in terms of $a_0$, $\tan \beta$
and two masses: either the bino mass $\tilde M_1$ and the average singlet charged slepton mass
$\bar m_R$, or the universal scalar and gaugino mass parameters, respectively $m_0$ and $M_{1/2}$. 
The upper limits on $|C_{\mu e}|$ from the present upper bound $\mathrm{BR}( \meg) <
10^{-11} $ are shown in fig.  \ref{F2-C12} for $\tan \beta=10$. The two
choices for $a_0$ are meant to give an idea of the theoretical
uncertainty:  a) $a_0=0$ (dotted lines) and b) $a_0 = m_0 + M_{1/2}$
(solid lines). For different choices of $\tan\beta$ and/or better
sensitivities than the present one, it is enough to rescale according
to (\ref{BR}). The same plot can be used to estimate the upper bounds
on $C_{\tau \mu}$ from $\tmg$ by taking into account the factor
$\mathrm{BR}(\tau \rightarrow \mu \bar{\nu} \nu) \approx 17\% $.

In spite of the much poorer determination of $\eps$ at the time, 
seesaw models of the present class were among the first analysed to get
predictions for $\meg$ and $\tmg$. It was pointed out that  for models 
whose charges allow for $O(1)$ couplings in $Y_\nu$, the 
predictions for $\meg$ are at the level of the present limits and the 
planned improvement \cite{exp-mueg-f} would test them 
\cite{BuchDelViss, SatoTobe} (as can also be inferred from fig. \ref{F2-C12}). 
The relevance of $\tmg$, $\teg$ was emphasized as well.  
Of course, such considerations surpass these particular models.
A general discussion of LFV decays for the three possible classes of
seesaw models, defined in terms of the dominance mechanism in
the neutrino masses, was presented in ref. \cite{LMS1}.

Actually, what characterizes the present class of  models is that all
$\nu^c$'s contributions to $m_{\nu}$ are on an equal-footing, as
expressed by eq. (\ref{specif}), so that

\bea
\label{Cmod}
C^k_{ij}&  = & h^*_{ki} h_{kj} ~\eps^{\delta _{i1} + \delta _{j1}} 
\eps^{2 (n_k +\ell+h_u)} \ln \frac{ M_\mathrm{Pl} }{ M_k } 
\nonumber \\
& =  & \frac{h^*_{ki} h_{kj}}{r_k d_@} \eps^{\delta _{i1} + 
\delta _{j1}} \frac{M_k}{M_@} \ln \frac{ M_\mathrm{Pl} }{ M_k } 
\ , \quad \forall k  ~.
\eea

\noindent Therefore, from the experimental limits on $|C_{ij}|$ one
gets directly bounds on the heaviest Majorana mass $M_3$ - unless, for
$M_3 \simeq M_2$, $ C^3_{ij}$ and $C^2_{ij}$ strongly interfere, 
which amounts to some tuning of the free parameters.  On the other
hand, the leptogenesis constraint on the lightest mass $M_1$, provides
a lower limit, $C_{ij} \gtrsim \eps^{\delta _{i1} + \delta _{j1}}
(10^{-2}-10^{-3})$, hence a lower bound on BR$(\ell_i \rightarrow \ell_j \gamma)$.

We first discuss in some detail the bound on $M_3$ which follows from $\meg$.  
The right panel of fig. \ref{F2-C12} shows the scatter plot of
$|C^3_{\mu e}|$ - eq. (\ref{Cmod}) - as a function of $M_3/M_@$ for the models fulfilling all
neutrino oscillation and leptogenesis constraints (of course similar plots apply 
also to $k=1,2$). Since, with the value of $\eps $ fixed as in section 
\ref{sec:OSC}, $|C^3_{\mu e}| = O( M_3/M_@)  $, the experimental upper 
bounds on $|C_{\mu e}|$ and $M_3/M_@$ are of the same order of magnitude:

\beq
\frac{M_3}{M_@} \lesssim  O( |C_{\mu e}|^{\mathrm{u.b.}} )~.
\label{M3-lim}
\eeq

\noindent These upper bounds are displayed in the left plot of fig. \ref{F2-C12}.  
With the present experimental sensitivity, they already provide 
meaningful limits on $M_3$, in spite of the uncertainties in the 
supersymmetric spectrum. For instance, with $\tan \beta = 10$,
$M_3/M_{@} \lesssim 0.3 (1)$ for the mass region below 
the orange (yellow, resp.) line, some two orders of magnitude below 
the gauge coupling unification scale! With the planned three orders of
magnitude improvement in BR$(\meg)$, it would become
$M_3  / M_@ \lesssim 10^{-2}$ in that mass regions,  which is 
close to  the $M_1$ range selected by leptogenesis, where $C_{\mu e} 
\gtrsim C^1_{\mu e} \approx (10^{-3} -10^{-4})$. Therefore, for this 
class of models, there are fair perspectives to detect a positive signal in 
$\meg$ searches. Alternatively, barring destructive interference among the
$C^k_{\mu e}$, the future experiments would rule out this whole class of 
models in the region of the parameter space where 
$|C_{\mu e}|^{\mathrm{u.b.}} \lesssim 10^{-4}$. 
For $\tan \beta \ge 20$, it actually coincides with the area below the red lines which also 
corresponds to a sizeable supersymmetric contribution to $(g-2)_{\mu}$
(see \cite{sleptonarium} and references therein).  
This clearly shows the power of the combined
constraints from $\meg$ and leptogenesis in this class of models.

\begin{figure}[!t]
\vskip .5 cm
\centerline{  \psfig{file=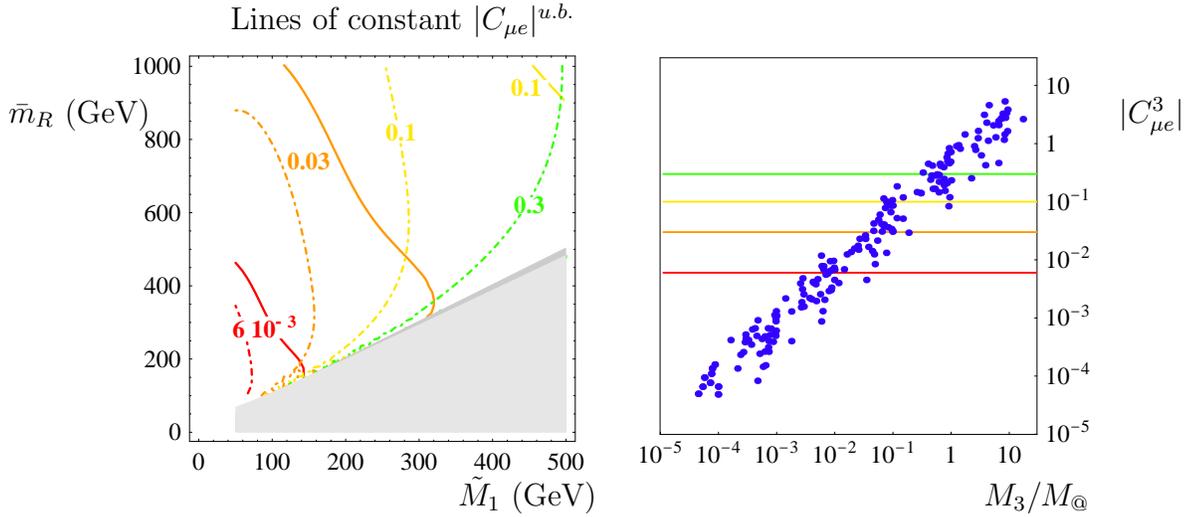,width=.9 \textwidth}
\put(-420, 145){$\bar m_R$ (GeV)}
\put(-250, 0){$\tilde M_1$ (GeV)}
\put(-340, 180){ Lines of constant $|C_{\mu e}|^{u.b.}$}
\put(0, 145){$|C^3_{\mu e}|$}
\put(-55, 0){ $M_3/M_@$} }
\caption{The left plot shows the upper bounds on $|C_{\mu e}|$ for
BR($\meg$) $\le 10^{-11}$ and $\tan \beta =10$. The plot, adapted
from that in \cite{IMsusy02}, includes both the chargino and neutralino
contributions and assumes mSugra with $a_0=0$ (dotted line) and  $a_0 =
m_0 + M_{1/2}$ (solid one).  The right plot shows the scatter plot of
$|C^3_{\mu e}|$, eq. (\ref{Cmod}), as a function of $M_3/M_@$ 
for the models fulfilling all neutrino oscillation and leptogenesis constraints,
with $M_1 < 10^{11}$ GeV. We take $M_3$ to span its allowed range, namely
$M_1 \le M_3 \le d_@ r_3$.}
\label{F2-C12}
\vskip .5cm
\end{figure}

As already noticed, the couplings to $e$ are reduced with respect to
$\mu$ and $\tau$ by a factor $O(\eps ) = O(m_\odot /m_@)$, so that
$C^k_{\mu e} = O(C^k_{\tau e}) = O(\eps) C^k_{\tau \mu}$.  This 
simple link to the $m_{\nu}$ pattern is a characteristic of the model.  In
particular,  the model predicts: BR($\meg $) : BR($\tmg$ ) : BR($\teg
$) = $\eps ^2 : .17 : .17 \eps ^2$, while the present experimental upper
bounds are in the ratios $10^{-5}  :  1  : 1 $, respectively.  Hence,
in this model, from the experimental limit on $\meg$, the other two LFV
transitions are predicted many orders of magnitude below their present
experimental limits.

\section{Lepton EDM} 

There is a rich literature about the constraints that lepton EDM searches 
put on the CPV pattern of the MSSM and its extensions \cite{nath-etc, romstru}, 
in particular, on the slepton mass matrices. It is useful to separate them 
into two kinds:  \textit{i)} lepton flavour conserving (FC) parameters, 
namely, the phase of the $\mu$-term  and phases in the flavour diagonal 
chirality flipping slepton masses, or A-terms; \textit{ii)} phases in the 
lepton flavour violating (FV) elements of the slepton mass matrices. Very 
restrictive conditions on some of those phases arise from the present limits 
on $d_e$ which are to be considerably strengthened in the near future 
\cite{exp-de-f}. The present limit on $d_\mu$ is not very significant but a 
new experiment \cite{exp-dmu-f} could bring relevant results. The first 
conclusion is that supersymmetric models must possess a mechanism to effectively 
inhibit CP phases in the soft terms at $M_{Pl}$. The second one is that the 
radiative corrections must keep the induced CP phases in the soft terms below 
the experimental limits. In this section we concentrate on the constraints on 
the radiative corrections arising from the Yukawa couplings of the supersymmetric 
seesaw models.

The main contributions to $d_{i}$'s of either  kind depend on the slepton mass 
parameters as follows:

\beq
\label{di}
d^{FC}_{i} = \frac{e^{3} \tilde M_{1} I_B}{32 \pi ^{2} |\mu|^2 \cos ^{2}\theta_W} 
m_{\ell_i} \mathrm{Im}(a_{i}) \, \qquad
d^{FV}_{i} = \frac{e^{3} \tilde M_{1} \mu \tan \beta I''_B }{32 \pi ^{2} |\mu|^{2} \cos^{2}\theta_W} 
\mathrm{Im}(\delta^{RR} m_{\ell} \delta^{LL} )_{ii}\,
\eeq

\noindent
in  the basis where $Y_\ell$ and $M_R$ are real and diagonal and with the 
FC  $A$-terms written as $A_i = a_i Y_{\ell_i}$. The functions $I_B$ and 
$I''_B$ depend on the slepton, chargino and neutralino masses as defined in 
ref. \cite{sleptonarium}. The first contributions to $d^{FV}_{i}$ are second 
order in the insertion approximation. Only the FV elements of the (hermitian) 
matrices $\delta^{RR}$, $\delta^{LL}$, those responsible for LFV decays, 
contribute in (\ref{di}). Hence $d^{FV}_e$ is linked to $\teg$ and $\meg$, 
$d^{FV}_\mu$  to $\tmg$ and $\meg$,  $d^{FV}_\tau$ to $\tmg$ and 
$\teg$ (see, \eg , \cite{sleptonarium} for a general discussion). Here, these 
links are analysed anew in the framework of the supersymmetric seesaw mechanism 
as one source of LFV and CPV.

We start with real $\mu$ and no CP or lepton flavour violations in the 
$A$-terms  at $M_{Pl}$. In particular,  $\mathrm{Im}a_{i} = 0$. Then,  CP 
violations are introduced through  the RG running only by the heavy neutrino 
couplings $Y_{\nu}$. The decoupling of right handed neutrinos yields an 
effective MSSM, with neutrino masses and CP phases in the slepton masses. 
As discussed in section \ref{sec:LFV}, the hermitian matrices $C^k$ appear 
at one loop and the  $| C^k_{ij}|$ are bounded by the $\ell_i \rightarrow 
\ell_j \gamma$ decays. However, since $C^k_{ii}$ is real, there is no CPV 
phase  to induce $d_i$. This means that the radiatively induced CP violations 
start at the two-loop level and, in the leading log development, come from the 
phases of LFV elements in the products $\mathrm{Im} ( \sum_j C^k_{ij}C^{k'}_{ji} ),\ k \neq k'$. 
By obtaining  the dependence of $d^{FC}_{i}$ and 
$d^{FV}_{i}$ on the $C^k_{ij}$ one makes more manifest how they are 
constrained by the LFV experiments. Let us first discuss these relations more
generally, before turning back to the class of models analysed here.

\subsection{General link between EDM and LFV}\label{sec:GenLink}

Recently,  the dominant two-loop contributions to Im$a_{i}$ and 
$\delta^{RR}$, $\delta^{LL}$ in the expressions of the supersymmetric 
seesaw radiatively induced EDMs have been calculated \cite{EDM-Ellis, 
EDM-Isa, FarPesk}.  Here, the results are rearranged to make clear the general 
connection between LFV and EDM,  by expressing $d_{i}$ in terms of the 
$C^k_{ij}$. The original results in \cite{EDM-Isa} and some detailed 
expressions are presented in App. A.

Assuming mSugra boundary conditions at $M_{Pl}$ for simplicity, the
result \cite{EDM-Ellis, EDM-Isa} for  $d^{FC}_i$ is obtained by replacing in (\ref{di}):

\beq
\mathrm{Im} (a_i) = \frac{8}{(4 \pi)^4} a_0 \sum_{k > k'} 
\frac{\ln (M_{k}/M_{k'})}{\ln (M_{Pl}/M_{k'})}~
 \mathrm{Im} ( \sum_{j \neq i} C^{k}_{ij}C^{*k'}_{ij} )
\label{ImA}
\end{equation}

\noindent
Notice that since $\mathrm{Im}(\sum_{i} a_i) = 0$  the FC contribution 
leads to the sum rule:

\begin{equation}
\label{SR}
\sum_{i} d_{i} / m_{i} = 0~.
\eeq

To obtain $d^{FV}_i$, one has to replace  in (\ref{di}) the leading two-loop 
result \cite{EDM-Isa, FarPesk} for the double insertion term, which is:

\begin{eqnarray}
\label{2deltas}
\mathrm{Im}(\delta^{RR} m_{\ell} \delta^{LL} )_{ii} &=& \frac{8}
{(4 \pi)^6}  \frac{(6 m_0^2 + 2 a_0^2) (6 m_0^2 + 3 a_0^2)}{\bar m_L^2 \bar m_R^2}
\frac{m^2_\tau}{m^2_t}m_{\ell_i} \tan^2 \beta ~I^{FV}_i ~,
\nonumber \\
I^{FV}_i &=& \sum_{k > k'} \widetilde{\ln}^{k}_{k'} ~\mathrm{Im} (
\sum_{j \neq i} C^{k}_{ij} \frac{m^2 _{\ell_{j}}}{m^2_\tau} C^{k'*}_{ij} ) ~, 
\end{eqnarray}

\noindent
where $\bar m_R$ and $\bar m_L$ are average masses of the sleptons of each chirality, 
and the logarithmic dependence on the masses $M_{k}$ represented by 
$\widetilde{\ln}^{k}_{k'}$ is defined in App. A. The sum rule (\ref{SR}) is not valid
as the FV contribution rather leads to the sum rule:
\beq
\sum_{i} m_i d_{i} = 0~.
\eeq

There are other three points to notice with respect to these expressions: 

\noindent \textit{i)} The EDM phase results from the (three different)
relative phases between $C^{k}_{ij}$ and $C^{k'}_{ij}$, namely between 
the $\nu^{c}_{k}$ and $\nu^{c}_{k'}$ contributions to $\ell_i \,\rightarrow \ell_j \gamma$.

\noindent \textsl{ii)} The phase in the quark CKM matrix is large 
and there is no known reason to think of smaller phases in the lepton 
sector, thus one expects the EDM phases to be generically large.

\noindent \textsl{iii)} As already discussed, $|C^k_{ij}|$ are constrained 
by the experimental limits on the LFV decays $\ell _i \rightarrow 
\ell _j\, \gamma$. However, at present the only significant bound from LFV 
decays is that on $|C_{\mu e}|$ from $\meg$. 

It has been shown in ref. \cite{EDM-Isa} that with (\ref{ImA}) and 
(\ref{2deltas}), $d_i^{FV} > d_i^{FC}$ for $\tan \beta \ge 10$. 
Indeed,  the FC contribution vanishes for $a_0=0$, while for $a_0 \sim m_0$, 
as $d_{i}^{FV} / d_{i}^{FC} \propto  (\tan \beta )^{3}$, the FV 
contribution dominates over the FC one for large $\tan \beta $ in spite of 
their different sparticle mass dependence. We refer to \cite{EDM-Isa} for 
general analytical and numerical studies of these seesaw mechanism as a source 
of lepton EDM. 

For those reasons, we mainly concentrate on $d_i^{FV}$. In fig. \ref{F2-de}, 
the limit on $I^{FV}_e$  obtained from the  bound $d_e \lesssim  
10^{-32}$ e\, cm is shown for $\tan \beta = 10$ and $a_0=0$ (dotted lines) 
or $a_0=m_0+M_{1/2}$ (solid ones). With the present bound the limits are 
$10^{5}$ larger and quite useless. Indeed, if the Yukawa couplings are at most  
$O(1)$ and the masses $M_{k}$ in the range selected by neutrino masses and 
leptogenesis,  one has $I^{FV}_i < O(10^{-3})$. Then, only for light sparticles 
and large $\tan \beta$ the present experimental bound on $d_e$ constrains 
$I^{FV}_e$ to be below this upper bound \cite{EDM-Isa}. 

\begin{figure}[!t]
\vskip .5 cm
\centerline{\psfig{file=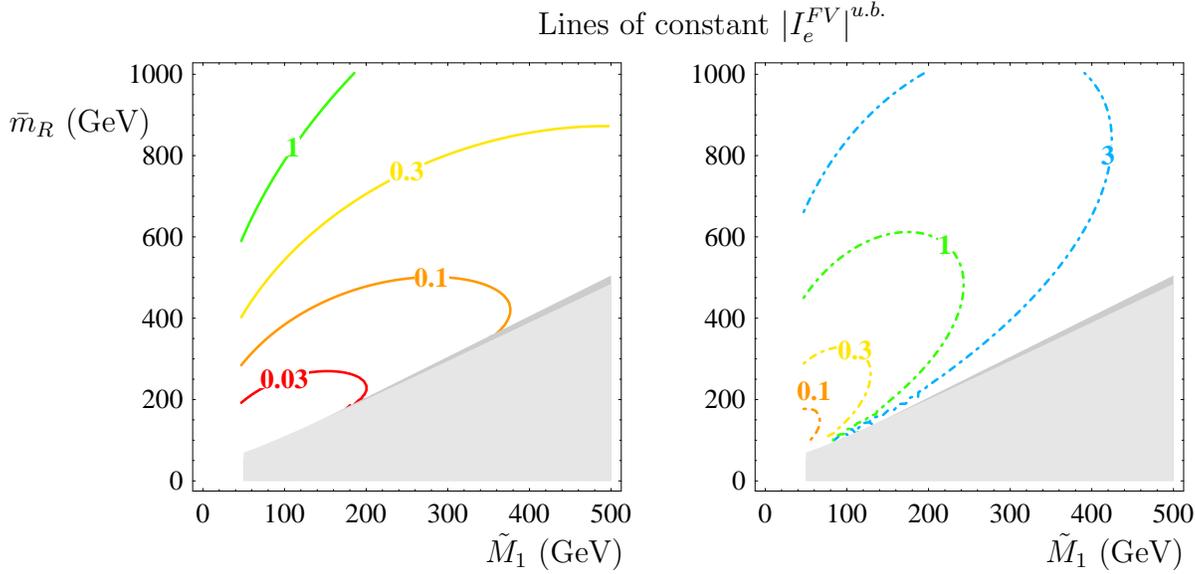,width= 1 \textwidth}
\put(-460, 163){$\bar m_R$ (GeV)}
\put(-280, 0){$\tilde M_1$ (GeV)}
\put(-68, 0){$\tilde M_1$ (GeV)}
\put(-260, 200){Lines of constant ${ |I^{FV}_e| }^{u.b.}$ } }
\caption{Upper limit on $|I^{FV}_e|$ from $ d_e \lesssim 10^{-32}$ e cm
with $\tan \beta =10$, $a_0= m_0 + M_{1/2}$ (solid line) and $a_0=0$ (dotted line).
For larger values of $\tan \beta$, multiply the upper bound for $(10/\tan\beta)^3$. }
\label{F2-de}
\vskip 0.5cm
\end{figure}

Finally, from the explicit expression for (\ref{2deltas}) ,
\begin{equation}
\label{I-1}
I^{FV}_e = \sum_{k > k'} \widetilde{ \ln^k_{k'}}  \left[ \mathrm{Im} 
( C^k_{e \tau} C^{k'*}_{e \tau}) + \frac{m^2 _\mu}{m^2_\tau} 
\mathrm{Im} (C^k_{e \mu} C^{k' *}_{e \mu} ) \right] 
\end{equation}

\noindent 
one sees that the limit on $d_{e}$ from LFV depends more on $\teg$ whose 
experimental upper bounds are still very far from  providing useful constraints. 
Therefore  this comparison is only possible in specific models and the class of 
models studied in this paper is particularly predictive in this respect as already 
discussed in section \ref{sec:LFV}.

\subsection{EDM in this class of  models} \label{EDMmodel}

In the class of models discussed in the previous sections, the  $C^k_{ij}$ are 
all related,  so that $C^k_{e \mu}$ and $C^k_{e \tau}$ are generically $O(\eps) C^k_{\mu \tau}$. 
Therefore they are all constrained by the most significant limit, 
that from BR$( \meg )$ on $C^k_{e \mu}$. The potentially most important 
contribution to $d_e$ comes from the first term of $I^{FV}_{e}$ in (\ref{I-1}) 
and is constrained by the decay $\teg $. Replacing this term with eq. 
(\ref{Cmod}) one obtains,

\beq
\label{appFV}
I^{FV}_e  =  \sum_{k > k'} \widetilde{ \ln^k_{k'}} \ln (M_{Pl}/M_{k}) \ln (M_{Pl}/M_{k'})
\frac{\eps^2 M_k M_{k'}}{M^2_@} ~\frac{\mathrm{Im}(h_{k1}^* h_{k3} h_{k'1} h_{k'3}^*)}{r_k r_k' d_@^2} 
\eeq

\noindent where the main contribution comes from $k=3,~ k'=2$, the other ones being 
reduced by factors $O(M_{1}/M_{2})$ and $O(M_{1}/M_{3})$, respectively. 
The $\meg$ decay giving already a limit on ${M_{3}}$, in principle, the 
experimental limits on the electron EDM and (\ref{appFV}) yields a 
bound on $M_{2}$. However, as we now turn to discuss, the seesaw 
contribution to the lepton EDM comes out too small to give any relevant 
limit in the present class of models.

The maximum values of (\ref{appFV}) correspond to $M_{2}=O(M_{3})$ and are roughly given by

\begin{equation}
I^{FV}_e  =  O(10^{-2}) \left( \frac{ M_{3}}{M_{@}}\ln{\frac{ M_{Pl}}{M_{3}}}\right)^{2} \ ,
\end{equation}

\noindent which is maximum near the maximum value of $M_{3}$. 
As shown in section \ref{sec:LFV}, for $\tan \beta \ge 10$ and sparticle
masses below the TeV, the present limit on $\meg$ requires $M_3 \lesssim M_{@}$, 
which implies $I_e^{FV} \lesssim 10^{-1}$.
The planned  improvement in $\meg$ searches, at the level 
of BR$(\meg)\lesssim 10^{-14}$, would imply $M_3/M_@ \lesssim 0.03$ 
in the same region of the MSSM parameter space and, correspondingly, 
$I_e^{FV} \lesssim  10^{-4}$, thus $d_e \lesssim 10^{-36}$ e\, cm.
Comparing these numbers with the plots in fig. \ref{F2-de}, one concludes that, even with a bound of 
$10^{-32}$ e\, cm on $d_{e}$, this radiative contribution is too low to be detected.
  
The plots in fig. \ref{F-de} show the dependence on $M_{2}$ for 
the models that satisfy all the experimental constraints of the previous sections. 
The vertical spreading of the points corresponds to the random overall phase. 
It is worth noting that the phases in $d_e$ and in $\varepsilon_1$ are 
uncorrelated. The former is generically proportional to $\mathrm{Im}(h_{31}^* 
h_{33} h_{21} h_{23}^*)$, while the latter rather depends on a combination of 
$\mathrm{Im}( h_{22} h_{12}^* + h_{23} h_{13}^*)$ and 
$\mathrm{Im}( h_{32} h_{12}^* + h_{33} h_{13}^*)$, as previously 
discussed in  section \ref{sec:LG}.

\begin{figure}[!t]
\vskip .5 cm
\centerline{\psfig{file=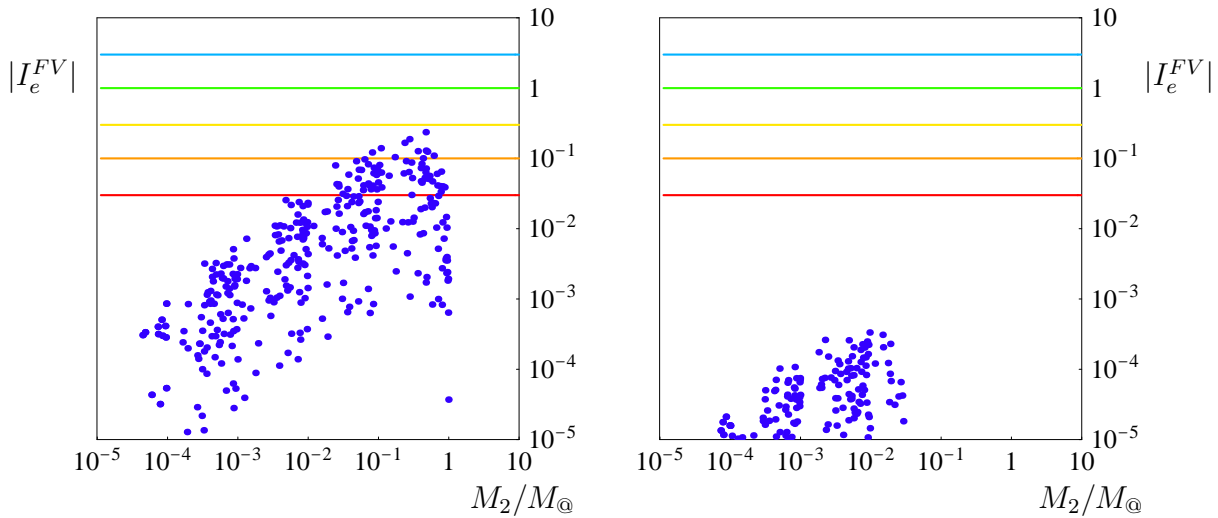,width= 1. \textwidth}
\put(-440, 160){$|I_e^{FV}|$}
\put(-265, 0){$M_2/M_@$}
\put(-10, 160){$|I_e^{FV}|$}
\put(-50, 0){$M_2/M_@$} }
\caption{ $I^{FV}_e$ as a function of $M_2/M_@$ for $M_3/M_@ =1$ and $0.03$ respectively,  
with randomly generated models that fulfill the neutrino oscillation and leptogenesis experimental constraints 
with $M_1 < 10^{11}$ GeV.}
\label{F-de}
\vskip 0.5cm
\end{figure}

Notice also the predictions 

\beq
d_{\mu} = O \left( \eps^{-2} \frac{ m_\mu }{ m_e }\right) d_e = O(6000) d_e~,
~~~d_{\tau} = O \left( \eps^{-2} \frac{m^2_\mu}{m_e m_\tau } \right) d_e= O(400) d_{e}~,
\eeq 

\noindent very different from the naive proportionality 
between lepton EDM and mass. Of course $d_\mu$ and $d_\tau$ are also predicted to be
extremely small.

\section{Decoupling the heavy neutrinos from $m_{@}$} 
\label{sec:DECOUPLING}

The fact that heavy Majorana masses end up pushed down by the experimental
limits on LFV is due to the main characteristic of the models studied until now:
as far as the heavy neutrinos commensurately contribute to the value of $m_{@}$,
the only way to reduce the LFV from the Yukawa couplings is to decrease their 
masses. The alternative is to decouple one (or two) heavy neutrino from
$m_{@}$. This is naturally done in the abelian flavour symmetry approach
by the introduction of the so-called analytic zeros, imposed by the
flavour symmetry and supersymmetry on the analytic superpotential. In the 
broken flavour symmetry phase, the zeros in the effective Yukawa interactions are 
filled  by the  diagonalisation  of the Kahler metric which becomes $\eps$ 
dependent. Apparently, their values are diminished by powers of $\eps$, 
and so might be their contribution to LFV decays. Actually it turns out that,
in the basis of the physical charged leptons, the couplings of the would-be 
decoupled state must be so much \textit{larger} than those in the coupled case 
that the (partial) decoupling option is already experimentally excluded.

In this section, we try and make the expressions more transparent by omitting
the $O(1)$ matrix elements defined and displayed in the previous ones. 
Correspondingly, we discuss only orders of magnitude and show the results in
this form. More detailed calculations and definitions are presented in 
Appendix B, including a brief account of the transformations to the basis 
chosen here with diagonal $M_{R}$ and $Y_{\ell}$. 
It is also shown that the hierarchical structure, \ie ,  the $\eps$ dependence of $m_{\nu}$ as 
defined by (\ref{m_nu}), is stable under the necessary field redefinitions.
From the discussion in section \ref{sec:OSC} we can already already
take $\ell_{2} = \ell_{3} =\ell =\ell_{1} - 1$, as in eq. (\ref{m_fit}),
as well the relations in (\ref{specif}) and (\ref{Mat}) below.

In order to decouple $\nu_{3}^{c}$ from $m_{@}$, we assume the charge 
$n_{3}$ to be large enough so that $n_{3} +  \ell + h_{u} < 0$. Since this 
charge imbalance cannot be compensated analytically, it implies that
$Y_{\nu\, 3i} = 0\ (i=2,3)$ in the basis where the abelian flavour charge
is diagonal. If one also chooses  $n_{3} +  \ell_{1} + h_{u} < 0$,
$\nu_{3}^{c}$ decouples from \textit{all} light neutrinos, in this basis, 
and the model becomes of the so-called $3\times2$ type, which we skip here
since it has received a lot of attention in the literature \cite{FGY}.  
As just explained, we are led to choose $\ell_{1} = \ell + 1$ from
(\ref{m_fit}). Hence $n_{3} +  \ell_{1} + h_{u} = 0$ and $Y_{\nu\, 31}
= O(1)$. Then, from  (\ref{Mat}), we get 

\begin{equation}
M_{3} = O(\eps^{-2} M_{@}) = O(10^{16})\mathrm{GeV} \ , \label{M3dec}
\end{equation}

\noindent close to the GUT scale. Since $M_{k}/M_{3}= O (\eps ^{2n_{k}-2n_{3}})$, 
one has in the basis of the flavour charge eigenstates:

\begin{equation}
Y_{\nu\, 31} = O(1)\ ,  \quad Y_{\nu\, 3j} = 0\quad (j=2,3)\ ,  \quad 
Y_{\nu\,kj} = O\left( \eps^{\delta_{j1}}\sqrt{\frac{M_{k}}{M_{@}} }
\right) \ (k=1,2) \  . \label{Yflav}
\end{equation}

In going to the physical basis where $M_{R}\ (Y_{\ell})$ is diagonal, 
there are two steps: first to put the Kahler metric in canonical form, then to diagonalise
$M_{R}\ (Y_{\ell}$, resp.). While in  section \ref{sec:MODEL} it has been remarked
that this does not change the structure of the matrices in the models
previously studied, here they fill the analytic zeros  $Y_{\nu\, 3j} = 0\ (j=2,3)$.
In this basis these couplings become (see Appendix B)

\begin{eqnarray}
Y_{\nu\, 3j} & \rightarrow & \sum_{k=1,2} O\left(\frac{\sqrt{M_3 M_{k}}}{M_{3}-M_k}\right) Y_{\nu\, k3} 
+ \eps \eta_{L1j} Y_{\nu\, 31} \ , \nonumber \\
& = &  O\left(\eps \frac{M_{2}}{M_{@}}  \right) + O(\eps) \qquad (i=2,3) 
\  ,  \label{Ydecoupl}
\end{eqnarray}

\noindent where the O(1) coefficients come from a small rotation and a small 
Kahler vielbein, the first (second) terms are associated to the
transformations of the right-handed neutrinos (left-handed charged leptons, resp.). 
Notice that these couplings are large, of $O(\eps)$, so large that the model is ruled out by
LFV decays. Indeed, keeping only the leading $\nu_{2}^{c}$ and $\nu_{3}^{c}$
contributions, one gets:

\begin{equation}
C_{\mu e} =  \eps~ O\left( \frac{M_{2}}{M_{@}}\right)\ln\frac{M_{Pl}}{M_{2}} 
+ \eps \left( O(1) + O\left(\frac{M_{2}}{M_{@}}\right)\right)
\ln\frac{M_{Pl}}{\eps^{-2}M_{@}}   = O(1)\ . \label{C12dec}
\end{equation}

\noindent This value is above its MSSM upper limit  $|C_{\mu e}|^{\mathrm{u.b.}},$
discussed in section \ref{sec:LFV}, even for quite heavy gauginos, 
${\tilde M}_{1} = O(\mathrm{TeV}).$ The only remaining possibility
is to assume $M_{2} = O (M_{@})$ and an (unnatural) cancellation between
the various $ O(1)$ terms that result. In this sense  models with decoupling are
\textsl{(i)} fixed up to $O(1)$ coefficients and  \textsl{(ii)} basically 
already excluded by experiments. Remember that planned experiments on $\meg$ will
reduce  $|C_{\mu e}|^{\mathrm{u.b.}}$ by more than an order of magnitude.

Finally, let us just comment on the remaining possibility of decoupling 
\textit{both} $\nu_{3}^{c}$ and $\nu_{2}^{c}$ from $m_{@}$,
namely from both $\nu_{2}$ and $\nu_{3}$, so that one neutrino
remains exactly massless while  $\nu_{1}^{c}$ dominates in $m_{@}$.
This could look interesting because,  the tuning of $\rho$ examined in section 
\ref{sec:OSC} might become natural \cite{King}. However, it is easy
to check that this configuration leads to $m_{\odot}/m_@ = O(\eps ^{2})$ and 
$\theta_{12} = O(\eps )$ requiring unnatural coefficients to fit the
oscillation data. Of course the same constraints as in the case of only one
decoupling from  $m_{@}$ also apply here and require a tuning between 
the contributions from $\nu_{3}^{c}$ and $\nu_{2}^{c}$ to $C_{e\mu}$,
which are both of $O(1)$. Altogether, this last case is even less attractive
than the previous one.

\section{Conclusions} 

In this paper we have emphasized the complementarity of some of the 
present and future experimental data for the 
determination  of the Dirac and Majorana mass matrices 
of the supersymmetric seesaw models. For this purpose, we have chosen 
the simplest abelian flavour models, in which the structure of the 
Dirac and Majorana masses are fixed, up to $O(1)$ parameters 
with arbitrary phases, by only 8 parameters: the 6 lepton flavour charges,
the small Froggatt-Nielsen parameter $\eps$ and the $B-L$ breaking scale.
We have shown that  the available data (basically the 
atmospheric and solar oscillation masses and angles, the baryon asymmetry
in the universe and the $\meg$ decay) are enough to fix the seesaw mass
patterns to a large extent. These models predict  the $\meg$ decay to
be within the  reach of the planned searches. They also
predict a very low contribution to $d_{e}$ which  is bounded in this class 
of models by the limits on  $\meg$ alone. In other words, the lepton
flavour abelian flavour models are consistent with experiments but only
a few choices of the parameters are possible and basically all the selected
models will be tested by the future experiments.  

The models possessing all these properties are theoretically well-defined
effective theories, the flavour symmetry is the simplest abelian one.  
The fact that  a few experimental data of very different nature can 
complement each other to constrain so much the models is certainly 
non-trivial. Type I leptogenesis always requires one relatively  light 
right-handed neutrino, but here it has to be at the upper side of the
allowed region. It is well-known that the $\meg$ experiments are testing  
the LFV effects in the slepton masses to the level of the radiative corrections, 
but here they immediately translate into a limit on a right handed neutrino 
mass eigenvalue and the whole series of LFV decays is predicted. Namely, 
the complementarity of the experimental constraints is disentangled into 
simple relations for the physical parameters.  

Of course these models are special in many respects. They predict a hierarchical
spectrum for the light neutrinos with $U_{e3}$ close to its present bound,
choosing one of the possible neutrino mass matrix patterns. They explains the large 
oscillation mixing angles by large mixings in the Dirac masses, favouring
large LFV. The ratio between the solar and the atmospheric mass differences
is directly related to a fundamental parameter, the scale of the flavour symmetry 
breaking, at the price of some tuning. Whenever all the right handed neutrinos couple to 
all left handed ones in the Yukawa matrix, they all contribute to the atmospheric
neutrino oscillations. If one right state  (not a mass eigenstate) decouples from the 
$\mu$ and the $\tau$ doublets, one Majorana mass eigenvalue can be quite large,
close to the GUT scale. With the similar decoupling of another state, one gets the
pattern corresponding to the dominance of the atmospheric oscillations by the 
lightest right handed neutrino, but $m_{\odot}/m_@$ and $\theta_{12}$ come out too small.

Thanks to supersymmetry, in the class of models examined here all these situations are 
naturally realized in the sense that they are ensured by the choice of the arbitrary charges.
The results are quite independent of the various $O(1)$ free parameters, 
meaning that the conclusions are technically robust. The dependence on the sparticle
masses is taken into account by analysing the results for a large range of values of 
the most important parameters (slepton and gaugino masses) . On this basis, our main 
conclusions here are: \textsl{(i)} models involving decoupling through supersymmetric
zeros are basically excluded and  \textsl{(ii)} the whole class of models will be
generically excluded in the near future, by the experimental bounds on the 
$\meg$ decay - once the constraints from the present data on neutrino oscillations 
and leptogenesis are imposed.  

General predictions of these models are the relations among the LFV charged
lepton decay rates, and the very low upper bounds on the contributions to 
$d_{e}$. Actually the latter are generically expected \cite{EDM-Ellis, EDM-Isa}  
if the seesaw couplings are the only source of CPV. Instead, if the couplings
of the GUT heavy colour triplets related to proton decay are also 
introduced, one can achieve much larger values of $d_{e}$ \cite{barhall, EDM-Isa}.
In $SU(5)$ GUT, with the $\bar 5$-plet and singlet charges fixed by the pure seesaw
contributions, the 10-plet ones by the quark masses and mixings, the limits on
$d_{e}$ are then predicted \cite{IeC} up to uncertainties in $O(1)$ coefficients in the seesaw
model, their phases, and the amounts of non-minimality in the GUT required
to fit the data (fermion masses and mixings, proton decay). 
This analysis of $d_e$ in a GUT framework is also supported by
the $O(m_\odot/m_@)$ value of $\eps$ obtained in the present paper, 
quite close indeed to the Cabibbo angle which is the natural 
choice in the quark sector.

\vskip 1cm

{\it Acknowledgements:} 
The authors thank P. Di Bari for useful observations.
The authors are grateful to the Dept. of Physics 
of "La Sapienza" University in Rome for the warm hospitality during the completion of this work.   
This work has been supported in part by the RTN European Program 
MRTN-CT-2004-503369.

\vskip 1cm


\section*{Appendix A}

The FC contribution of $A_i \equiv a_i Y_{\ell_i}$ has been calculated 
in refs. \cite{EDM-Ellis} and \cite{EDM-Isa} at  lowest order\footnote
{The coefficients in the papers of \cite{EDM-Ellis} numerically disagree 
with each other and with the result in ref. \cite{EDM-Isa}. The differences are
irrelevant for the qualitative conclusions herein. Ref. \cite{FarPesk} find, 
on the contrary, a smaller FC contribution. }. In order to treat the various 
thresholds in a compact way, it is convenient to introduce the function defined as 
$\ln^{a}_{b} = \ln(M_a/M_b) $ where $a,b=1,2,3,4$ with the  identification 
$M_4=M_{Pl}$, as well as the matrices $N_a = Y^\dagger_\nu P_a Y_\nu$ 
with $P_3 =\mathbb{I}$, $P_2 =$diag$(1,1,0)$, $P_1  =$diag$(1,0,0)$, and it 
is understood that the Yukawas are taken at $M_{Pl}$.
Then \cite{EDM-Isa}:

\beq
\label{ImA }
\mathrm{Im}(a_i)= \frac{8}{(4 \pi)^4} a_0  I^{FC}_i  \quad , \quad \quad
I^{FC}_i = \sum_{a > b} \ln^{a+1}_{a} \ln^{b+1}_b  \mathrm{Im} (N_a N_b )_{ii}~
\eeq

\noindent where the sum is over $a,b=1,2,3$.

The contribution  from two insertions of FV slepton mass matrix
elements obtained in \cite{EDM-Isa, FarPesk} reads:

\bea
\label{}
\mathrm{Im}(\delta^{RR} m_{\ell} \delta^{LL} )_{ii} =
\frac{8}{(4 \pi)^6} m_{\ell_i} \frac{m^2_\tau}{m^2_t}\tan^2 \beta  \frac{(6 m_0^2 + 2 a_0^2) 
(6 m_0^2 + 3 a_0^2)}{\bar m_L^2 \bar m_R^2} I^{FV}_i ~~ , \nonumber \\
I^{FV}_i= \sum_{a > b} \ln^{a+1}_a \ln^{b+1}_b ( \ln^{a+1}_a + \ln^{b+1}_b )  
\mathrm{Im} (N_a \frac{m^2 _{\ell}}{m^2_\tau} N_b )_{ii} ~~~~~~~~~
\eea

\noindent where $\bar m_R$ and $\bar m_L$ are average R and L slepton masses at low energy, 
and $Y _{\ell} \simeq  m_{\ell} \tan \beta /m_t $ has been substituted. 

The FC contribution vanishes for $a_0=0$. On the other hand, for $a_0 \sim m_0$, 
it turns out that the ratio $d_i^{FV}/d_i^{FC}$ is well approximated by the simple 
rule \cite{EDM-Isa}

\beq
\frac{d_i^{FV}}{d_i^{FC}} \approx 0.3 ~\frac{\ln^{Pl}_2}{10} ~\left(\frac{\tan \beta}{10}\right)^3 ~,
\eeq

\noindent where $\mu$ has been fixed by electroweak symmetry breaking. Thus, 
the FV contribution is in general favoured by a factor $(\tan \beta / 10)^3 $. 
A detailed comparison of these two contributions is in \cite{EDM-Isa}, 
together with  limits on $I^{FV,FC}_i$ from the present and planned sensitivity 
to $d_e$ and $d_\mu$.

The above expressions for $I^{FV}$ and $I^{FC}$ can be rewritten in terms 
of the $C$'s related to LFV as follows:

\bea
I^{FV}_i= \sum_{a > b} I^{(ab)FV}_i~,\qquad \qquad
I^{(ab)FV}_i = \widetilde{\ln^a_b}~ \mathrm{Im} ( C^a \frac{m^2 _\ell} {m^2_\tau} C^b )_{ii}~,\\
\widetilde{ \ln^3_2 }= \ln^3_2~, \qquad 
\widetilde{ \ln^3_1 }= \ln^3_1 - 2 \frac{\ln^3_2~ \ln^2_1 }{ \ln^{Pl}_1 }~,\qquad 
\widetilde{ \ln^2_1 }= \ln^2_1 \left(1 - 2 \frac{\ln^{Pl}_3 ~\ln^3_{2} }{ \ln^{Pl}_1 ~\ln^{Pl}_2 }\right) ~.
\eea

\noindent Hence, explicitely:

\bea
I^{FV}_1 &= & \sum_{a > b} \widetilde{ \ln^a_b}  
\left[ \mathrm{Im} ( C^a_{13} C^{b *}_{13}) + \frac{m^2 _\mu}{m^2_\tau} 
\mathrm{Im} (C^a_{12} C^{b *}_{12} ) \right] \\
I^{FV}_2 &= & \sum_{a > b} \widetilde{ \ln^a_b}  
\left[ \mathrm{Im} ( C^a_{23} C^{b *}_{23}) - \frac{m^2 _e}{m^2_\tau} 
\mathrm{Im}(C^a_{12} C^{b *}_{12} ) \right] \\
I^{FV}_3 &= & \sum_{a > b} \widetilde{ \ln^a_b}  \left[ - \frac{m^2 _\mu}{m^2_\tau}
\mathrm{Im} ( C^a_{23} C^{b *}_{23}) - \frac{m^2 _e}{m^2_\tau} 
\mathrm{Im}(C^a_{13} C^{b *}_{13} ) \right]
\eea

\noindent On the other hand, for the FC contribution,

\beq
I^{FC}_i=\sum_{a > b} I^{(ab)FC}_i~,\qquad 
I^{(ab)FC}_i = \frac{\ln^a_b}{\ln^{Pl}_b} \mathrm{Im} ( C^a  C^b )_{ii}~.
\eeq

\noindent Explicitely: 

\bea
I^{FC}_1 & = & \sum_{a > b} \frac{\ln^a_b}{\ln^{Pl}_b} \left[ \mathrm{Im} 
( C^a_{13} C^{b *}_{13}) + \mathrm{Im}(C^a_{12} C^{b *}_{12} ) \right] \\
I^{FC}_2 & = & \sum_{a > b} \frac{\ln^a_b}{\ln^{Pl}_b}  \left[ \mathrm{Im} 
( C^a_{23} C^{b *}_{23}) - \mathrm{Im}(C^a_{12} C^{b *}_{12} ) \right] \\
I^{FC}_3 & = & \sum_{a > b} \frac{\ln^a_b}{\ln^{Pl}_b}  \left[ - \mathrm{Im} 
( C^a_{23} C^{b *}_{23}) - \mathrm{Im}(C^a_{13} C^{b *}_{13} ) \right]
\eea

\noindent It is then manifest that $d_e$ is linked to $\teg$ and $\meg$, $d_\mu$ to 
$\tmg$ and $\meg$, and $d_\tau$ to $\tmg$ and $\teg$.

\section*{Appendix B}

This appendix contains some details of the results discussed in section 
(\ref{sec:DECOUPLING}), namely, when a heavy neutrino partially
decouples from the $\mu$ and the $\tau $. In the basis where $U(1)_{F}$ is
diagonal this is naturally implemented by supersymmetry zeros in the 
$Y_{\nu}$ matrix. In the more convenient basis where  $Y_{\ell}$ 
and $M_{R}$ are diagonal, these zeros are filled by: 
\textit{i)} the vielbein redefinitions of both the light and heavy states \cite{nirseiberg}, 
and \textit{ii)} the unitary transformations - that remain free in the vielbein - to go into this 
basis, in this order. In the framework of this paper, where generic $O(1)$ 
coefficients are free parameters and the calculations are carried on at
lowest order, these tranformations commute. Therefore, one has
to consider the effect of the tranformations:

\begin{eqnarray}
M_{R} & \rightarrow & V_{R}^{T} \xi ^{T}_{R} M_{R}  \xi_{R} V_{R} = 
\hat{M} \ , \nonumber \\
Y^{\dagger }_{\ell} Y_{\ell} & \rightarrow  & V^{\dagger }_{L} \xi_{L} ^{\dagger} 
Y^{\dagger }_{\ell} Y_{\ell}  \xi_{L} V_{L} = \hat{Y}^{2}_{\ell}\ , \nonumber \\
Y_{\nu} & \rightarrow & V_{R}^{T} \xi ^{T}_{R} Y_{\nu} \xi_{L} V_{L} =
\hat{Y}_{\nu}^{2} \ ,
\label{canonic}
\end{eqnarray}

\noindent where $\xi_{L,R}$ satisfy $(\xi \xi^{\dag })^{i\bar{j}} = 
K^{i\bar{j}}$, $K_{\bar{i}j}$ is the Kahler metric, $V_{R,L}$ are unitary, $\hat{M}$ 
and $\hat{Y}_{\ell}$ are diagonal. The metric is hermitean and invariant under $U(1)_{F}$, 
so that the transformation (\ref{canonic}) can be generally written at the lowest order as follows,

\begin{equation}
(\xi_{R} V_{R})_{ij} = \delta_{ij} + \eta_{Rij} \eps^{| n_{i} -n_{j} |} \ , 
\qquad  
(\xi_{L} V_{L})_{ij} = \delta_{ij} + \eta_{Lij} \eps^{|\ell_{i} -\ell_{j} |}
\ , \label{transf}
\end{equation}

\noindent where $\eta _{R,L}$  are $O(1)$ matrices whose hermitean parts 
bring the Kahler metric to the canonic form and whose anti-hermitean ones
diagonalise $M_{R}$ and $Y_{\ell}$, respectively. 
Of course, the non-diagonal terms are not restricted by analyticity and there 
are no vanishing matrix elements, in general. When applied to $Y_{\nu}$ this 
transformation fills the possible analytic zeros already at the first order, as
follows:

\begin{equation}
Y_{\nu ij} \rightarrow  \hat{Y}_{\nu ij} \approx Y_{\nu ij} + 
\sum_k Y_{\nu kj} \eta_{Rki} \eps^{| n_{i} -n_{k} |} + 
\sum_k Y_{\nu ik} \eta_{Lkj} \eps^{| \ell_{j} - \ell_{k} |}
 \ .    
\label{Yhat}
\end{equation}

\noindent If $n_{i}$ and $\ell{j}$ are such that $Y_{\nu ij} $ is a supersymmetric zero, 
it becomes ($n_{i}, \ell{j}$ decreasing with $i,j$):

\begin{equation}
\hat{Y}_{\nu ij} \approx
 \sum_{k<i} \eta_{Rki} h_{kj} \eps ^{2n_{k}-n_{i}+\ell _{j}+h_{u}}
+ \sum_{k<j} \eta_{Lkj} h_{ik} \eps ^{2\ell_{k}-\ell _{j}+n_{i}+h_{u}}~.
\label{0fill}
\end{equation}

Notice that, for hierarchical $M_{R}$, the rotation angles of $V_{R}$ are:

\begin{equation}
| \theta_{ki}|   \approx \left| \frac{r_{ik}}{\sqrt{r_{ii}r_{kk}}}
\frac{\sqrt{M_{i}M_{k}}} {M_{i} - M_{k}}
\right| \label{generat}
\end{equation}

\noindent and the vielbein elements are of the same orders of magnitude.

A few comments are in order here. When the heavy neutrinos are integrated out
to yield the effective $m_{\nu}$, the integration preserves analyticity as
it does not involve the Kahler metric and the results could dependent only on
the transformation $\xi_{L} V_{L}$ in (\ref{canonic}). In the 
basis of the physical charged leptons one has at the lowest orders:

\begin{eqnarray}
\hat{m}_{\nu ij} & = & (V^{T}_{L}\xi ^{T}_{L} m_{\nu}\xi _{L}V_{L})_{ij}
\nonumber \\
& \approx & m_{\nu ij} + \eta_{Lki} \eps^{| \ell_{i} - \ell_{k} |} m_{\nu kj}  
+ m_{\nu ik}\eta_{Lkj} \eps^{| \ell_{j} - \ell_{k} |} 
\label{hatmnu}
\end{eqnarray}

\noindent with $m_{\nu}$ given by eq. (\ref{m_nu}). It is easily checked that 
$\hat{m}_{\nu}$ has the same hierarchies in terms of $\eps $ as $m_{\nu}$,
with the $O(1)$ coefficients reshuffled by $\xi_{L}V_{L}$. Instead, in the calculation 
of the effects of the RG evolution in LFV and CPV, the metric dependence is
explicit in the results, even though they remain independent of $V_{R}$.
For leptogenesis, the lepton asymmetry clearly depends on which is the lightest 
heavy neutrino and the whole transformation $\xi_{R} V_{R}$ is needed.
Therefore it is convenient to work in the physical basis of the heavy
neutrinos and to express all the quantities in this basis as done in section 
\ref{sec:DECOUPLING}.

\vskip 1cm


\end{document}